\documentclass[12pt,a4paper]{article}
\usepackage{mathrsfs}
\usepackage{epsfig}
\usepackage{slashed}
\usepackage{amssymb}
\usepackage{amsmath}
\usepackage{amsthm}
\begin{document}
\title{Large Extra Dimension effects through Light-by-Light Scattering at the CERN LHC}
\author{Hao Sun\footnote{haosun@mail.ustc.edu.cn \hspace{0.2cm} haosun@dlut.edu.cn} \\
{\small Institute of Theoretical Physics, School of Physics $\&$ Optoelectronic Technology,} \\
{\small Dalian University of Technology, Dalian 116024, P.R.China}}
\date{}
\maketitle

\begin{abstract}

Observing light-by-light scattering at the Large Hadron Collider (LHC) has received quite
some attention and it is believed to be a clean and sensitive channel to possible new physics.
In this paper, we study the diphoton production at the LHC
via the process $\rm pp\rightarrow p\gamma\gamma p\rightarrow p\gamma\gamma p$
through graviton exchange in the Large Extra Dimension (LED) model.
Typically, when we do the background analysis, we also
study the Double Pomeron Exchange (DPE) of $\gamma\gamma$ production.
We compare its production in the quark-quark collision mode to the gluon-gluon collision mode
and find that contributions from the gluon-gluon collision mode are comparable to the
quark-quark one. Our result shows, for extra dimension $\delta=4$,
with an integrated luminosity $\rm {\cal L} = 200 fb^{-1}$ at the 14 TeV LHC, that diphoton
production through graviton exchange can probe the LED effects up to the scale
$\rm M_S=5.06 (4.51, 5.11) TeV$ for the forward detector acceptance $\xi_1 (\xi_2, \xi_3)$,
respectively, where $0.0015<\xi_1<0.5$, $0.1<\xi_2<0.5$ and $0.0015<\xi_3<0.15$.

\vspace{-0.5cm} \vspace{2.0cm} \noindent
 {\bf Keywords}: Large Extra Dimension, Forward Detector, Large Hadron Collider  \\
 {\bf PACS numbers}: 12.60.-i, 14.70.Bh
\end{abstract}

\newpage
\section{Introduction}

The Large Hadron Collider (LHC) at CERN generates high energetic proton-proton
($\rm pp$) collisions with a luminosity of $\rm {\cal L}=10^{34}cm^{-2}s^{-1}$
and provides the opportunity to study very high energy physics.
At such a high energy, most attention is usually paid to the central rapidity
region where most of the particles are produced and where most of the
high $\rm p_T$ signal of new physics is expected. Indeed, the CDF collaboration
has already observed such a kind of interesting phenomenon
including the exclusive lepton pairs production \cite{ppllpp1}\cite{ppllpp2},
photon photon production \cite{pprrpp}, dijet production \cite{ppjjpp} and
charmonium ($J/\psi$) meson photoproduction \cite{ppJPHIpp}, etc.
Now, both ATLAS and CMS collaborations have programs of forward physics
which are devoted to studies of high rapidity regions
with extra updated detectors located in a place nearly 100-400m
close to the interaction point \cite{FDs1,FDs2}.
Technical details of the ATLAS Forward Physics (AFP) projects
can be found, for example, in Refs.\cite{AFP,AFP1}.
The physics program of this new instrumentation covers interesting topics like
elastic scattering, diffraction, low-x QCD,
Central Exclusive Production (CEP) and the photon-photon ($\gamma\gamma$) interaction,
the last two being the main motivation for the AFP project.

CEP is a class of processes in which the two interacting protons are not destroyed
during the collision but survive into the final state with additional particle
states. Protons of this kind are named intact or forward protons.
This is a rare interaction and can take place through strong effects like Pomeron exchange
or electromagnetic effects, i.e., via photon exchange.
The kinematics of a forward proton is often described by means of the reduced energy loss $\xi$:
\begin{eqnarray}
 \xi=\rm \frac{\Delta E}{E}=\frac{E-E'}{E}
\end{eqnarray}
where E is the initial energy of the beam, $\rm E'$ is the energy after the
interaction and $\rm \Delta E$ is the energy that the proton lost in the interaction.
For CEP, the simple approximate relation between
the reduced energy losses of both protons ($\xi_1$ and $\xi_2$ )
and the mass of the centrally produced system M is
\begin{eqnarray}
\rm M^2=s \xi_1 \xi_2
\end{eqnarray}
where $\rm s=4 E^2$ is the square of the centre-of-mass energy.

The simplest exclusive production is due to the exchange of two photons:
$\rm pp\rightarrow p\gamma\gamma p\rightarrow pXp$ where X is the centrally produced system.
In such production at the LHC, the invariant mass of the photons can span up to 1 TeV scales,
high enough to reach scales of possible new physics.
In addition, the production of these processes mainly through the QED mechanisms which
are well understood and their predictions have a very small uncertainty.
These make the two-photon exchange physics particularly interesting.
Therefore, we can use this kind of production mechanism
to determine the luminosity at the LHC precisely\cite{LHCluminosityDeter},
to study the interaction of electroweak bosons with over-constrained kinematics, to test
the standard model (SM) at high energies or to study the new production channels
that could appear, i.e., SUSY\cite{SUSYprrp1, SUSYprrp2},
anomalous gauge couplings\cite{anoWWr1, anoWWr2, anoVVV, anoWWrr, anoZZZ, anoVVVV},
unparticle\cite{unparticle} and extra dimensions\cite{EDpllp1, EDprrp2}, etc.

Observing light-by-light scattering at the Large Hadron Collider (LHC)
has received quite some attention\cite{lightlight}
and it is believed a clean and sensitive channel to possible new physics.
In this paper, we study the diphoton signal from graviton exchange
in the Large Extra Dimension (LED) model via the main reaction
$\rm pp\rightarrow p\gamma\gamma p\rightarrow p G_{kk} p\rightarrow p\gamma\gamma p$
where $\rm G_{KK}$ is the KK graviton in LED.
A similar study has been performed in Ref.\cite{EDprrp2} where the authors study the
diphoton signal in both the LED and Randall-Sundrum (RS) models
and take $\gamma\gamma$ SM production as the corresponding background.
In our study we also consider the Double Pomeron Exchange (DPE) production
of the diphoton and present their cross section dependence on the energy loss of the proton $\xi$
and compare its production separately in the quark-quark collision mode
to the gluon-gluon collision mode. Our paper is organized as follows:
we build the calculation framework in Sect. 2 including a brief
introduction to the central exclusive diphoton production and Equivalent Photon Approximation (EPA),
the general diphoton exchange process cross section and a brief introduction to the LED model.
Section 3 is arranged to present the input parameters and numerical results of our study.
Typically we present a discussion of DPE diphoton production.
Finally we summarize our conclusions in the last section.

\section{Calculation Framework}

\subsection{Central Diphoton Exchange at the LHC and EPA}

\begin{figure}[hbtp]
\centering
\includegraphics[scale=0.8]{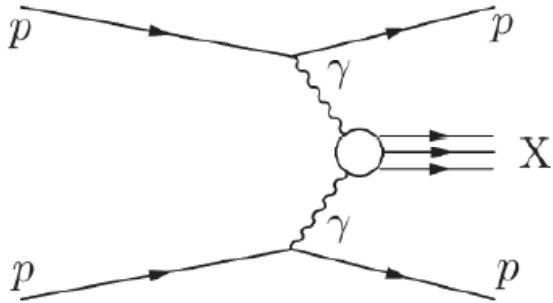}
\caption{\label{rrexclusive}
A generic diagram for the two photon exclusive production,
$\rm pp \rightarrow p\gamma\gamma p \rightarrow pXp$ at the CERN LHC.
Two incoming protons are scattered quasi elastically at very small angles.
The produced system X can be detected in the central detectors.}
\end{figure}

A generic diagram for the two photon central exclusive production
$\rm pp\rightarrow p\gamma\gamma p\rightarrow pXp$ is presented in Fig.\ref{rrexclusive}.
When the invariant mass of system X ($\rm M_{X}$) produced in the two-photon process is not too small,
we can factorize the amplitudes for the interaction $\rm \gamma\gamma\rightarrow X$
and introduce equivalent photon fluxes which play a similar role to the parton density functions
for the hadron interactions. This is indeed the basis of the Equivalent Photon Approximation (EPA)\cite{EPA},
and it allows us to calculate the proton cross section as the photon cross section plus two
equivalent photon fluxes. Two quasi-real photons emitted by each proton interact with each other
to produce X: $\rm \gamma\gamma\rightarrow X$.
Deflected protons and their energy loss will be detected by the forward detectors
while the X system will go to central detector.
In our case X system is the final state diphoton production.
A final state photon with rapidity $|\eta| < 2.5$ and some transverse momentum $\rm p_T > (5 - 10) GeV$
will be detected by the central detector.
The initial exchanged photons emitted with small angles by the protons
show a spectrum of virtuality $\rm Q^2$ and the energy $\rm E_\gamma$.
This is described by the EPA mentioned above which differs from the point-like electron (positron) case
by taking care of the electromagnetic form factors in the equivalent
$\gamma$ spectrum and effective $\gamma$ luminosity:
\begin{equation}\label{EPAformula}
\rm \frac{dN_\gamma}{dE_\gamma dQ^2}=\rm
\frac{\alpha}{\pi}\frac{1}{E_\gamma Q^2}\left[(1-\frac{E_\gamma}{E})(1-\frac{Q^2_{min}}{Q^2})F_E
 + \frac{E^2_\gamma}{2 E^2}F_M \right]
\end{equation}
with
\begin{eqnarray} \nonumber
\rm Q^2_{min}=( \frac{M^2_{inv} E}{E-E_{\gamma}} -M^2_P )\frac{E_{\gamma}}{E},
 ~~~~ F_E= \frac{4 M^2_p G^2_E + Q^2 G^2_M}{4 M^2_p +Q^2}, \\
\rm G^2_E=\frac{G^2_M}{\mu^2_p}=(1+\frac{Q^2}{Q^2_0})^{-4}, ~~~~F_M=G^2_M, ~~~~Q^2_0=0.71 GeV^2 ,
\end{eqnarray}
where $\rm \alpha$ is the fine-structure constant, E is the energy of the incoming proton beam.
which is related to the quasi-real photon energy by $\rm E_\gamma=\xi E$.
$\rm M_p$ is the mass of the proton and $\rm M_{inv}$ is the invariant mass of the final state.
$\rm \mu^2_p$ = 7.78 is the magnetic moment of the proton.
$\rm F_E$ and $\rm F_M$ are functions of the electric and magnetic form factors
given in the dipole approximation.

We denote the general diphoton exchange processes at the LHC as
\begin{eqnarray}
\rm pp \rightarrow p\gamma\gamma p\rightarrow p +ijk...+ p
\end{eqnarray}
with i, j, k, ... the centrally produced final state particles.
The partonic cross section $\hat\sigma_{\gamma\gamma\rightarrow ijk...}$
for the subprocess $\rm \gamma\gamma\rightarrow ijk...$ should be integrated over the photon spectrum
and we can obtain the total cross section
\begin{eqnarray} \nonumber
\rm \sigma_{pp} &=&
\rm \int^{\sqrt{s}}_{\omega_0} \hat\sigma_{\gamma\gamma\rightarrow ijk...} (\omega_{\gamma\gamma})
               \frac{dL^{\gamma\gamma}}{d\omega_{\gamma\gamma}}
               d\omega_{\gamma\gamma}  \\\nonumber
&=&\rm \int^{\sqrt{s}}_{\omega_0} \hat\sigma_{\gamma\gamma\rightarrow jj} (\omega_{\gamma\gamma})
  \int^1_{ \omega^2_{\gamma\gamma}/s } \frac{2\omega_{\gamma\gamma}}{x s}
  \int^{Q^2_{max}}_{Q^2_{1,min}} f_1(x,Q^2_1)  \\
& &\rm \int^{Q^2_{max}}_{Q^2_{2,min}}
  f_2(\frac{\omega^2_{\gamma\gamma}}{x s},Q^2_2)
 dQ^2_1 dQ^2_2 dx d\omega_{\gamma\gamma},
\end{eqnarray}
where $\rm \sigma_{pp}$ is the total photon-photon cross section, $\omega_{\gamma\gamma}$
is the two-photon center-of-mass (c.m.s.) energy, or the invariant mass of the produced system M,
$\rm s=4 E^2$, $\omega_0$ is some initial energy
and $\rm Q^2_{max}=2 GeV^2$ is the maximum virtuality. $\rm f=\frac{dN}{dE_\gamma dQ^2}$
is the $\rm Q^2$ dependent relative luminosity spectrum presented in Eq.(\ref{EPAformula}).
Consider $\rm E_\gamma=\xi E$ is some fraction of the total beam energy,
$\rm M^2=\omega^2=\xi_1 \xi_2 s= 4 \xi_1\xi_2 E^2$
and the forward detector acceptance satisfies $\rm \xi_{min}\leq\xi\leq\xi_{max}$,
we can also write the $\xi$-dependent cross section as
\begin{eqnarray}\label{totcrosssection}
\rm \sigma_{pp} = \rm \int^1_{\tau_0}
 \int^{\xi_{max}}_{\xi_0}  \int^{Q^2_{max}}_{Q^2_{1,min}}  \int^{Q^2_{max}}_{Q^2_{2,min}}
  \hat\sigma_{\gamma\gamma\rightarrow ijk...} (\xi_1 \xi_2 s)  f_1(\xi_1,Q^2_1) f_2(\xi_2,Q^2_2)
 dQ^2_1 dQ^2_2 d\xi_1 d\xi_2,
\end{eqnarray}
with $\rm \tau_0=M^2_{inv}/s$, $\rm M_{inv}$ is the invariant mass of the X system.
$\rm \xi_0=Max(\xi_{min},\tau_0/\xi_{max})$ .
The cross section of the subprocess $\hat\sigma$ can be written as
\begin{eqnarray}\label{partoiccrosssection}
 \hat\sigma = \rm
\int \frac{1}{avgfac} \frac{|{\cal M}_n ( \xi_1 \xi_2 s )|^2}{2 \hat s (2 \pi)^{3n-4}} d\Phi_n
\end{eqnarray}
where $\rm \frac{1}{avgfac}$ is the factor of the spin-average factor, the color-average factor,
and the identical particle factor (if there is any). $\rm |{\cal M}_n|^2$
presents the squared n-particle matrix element
and is divided by the flux factor $\rm [2 \hat s (2 \pi)^{3n-4}]$. The n-body phase space differential
$\rm d\Phi_n$ and its integral $\rm \Phi_n$ depend only on $\rm \hat s$ and particle masses $\rm m_i$
due to Lorentz invariance:
\begin{eqnarray} \nonumber
\rm \Phi_n(\hat s, m_1, m_2, ..., m_n) &=&\rm \int d\Phi_n(\hat s, m_1, m_2,..., m_n) \\
&=&\rm \int \delta^4((p_i+p_j)-\sum^{n}_{k=1}p_k) \prod^{n}_{k=1}d^4 p_k
\delta(p^2_k-m^2_k) \Theta (p^0_{k})
\end{eqnarray}
with i and j denoting the incident particles and k running over all outgoing particles.

\subsection{LED and Light-Light Scattering through Graviton Exchange}

The hierarchy problem strongly suggests the existence of new physics beyond the SM
at TeV scale\cite{Antoniadis}. The idea that there exist extra dimensions (ED)
which was first proposed by Arkani-Hamed, Dimopoulos,
and Dvali\cite{ADD}, might provide a solution to this problem.
They proposed a scenario in which the SM field is constrained to the
common 3+1 space-time dimensions ("brane"), while gravity is free
to propagate throughout a larger multidimensional space $\rm D=\delta+4$
("bulk"). The picture of a massless graviton propagating in D
dimensions is equal to the picture that numerous massive
Kaluza-Klein (KK) gravitons propagate in four dimensions. The
fundamental Planck scale $\rm M_S$ is related to the Planck mass scale
$\rm M_{Pl}=G_N^{-1/2}=1.22\times10^{19}~{\rm GeV}$ according to the
formula $\rm M^2_{Pl}=8\pi M^{\delta+2}_{S} R^\delta$ , where R and
$\delta$ are the size and number of the extra dimensions,
respectively. If R is large enough to make $\rm M_S$ on the order of
the electroweak symmetry breaking scale ($\rm \sim 1~ {\rm TeV}$), the
hierarchy problem will be naturally solved, so this extra dimension
model is called the large extra dimension model (LED) or the ADD
model. Postulating $\rm M_S$ to be 1 TeV, we get $\rm R\sim 10^{13}~{\rm
cm}$ for $\delta=1$, which is obviously ruled out since it would
modify Newton's law of gravity at solar-system distances; and we get
$\rm R\sim 1~{\rm mm}$ for $\delta=2$, which is also ruled out by
torsion-balance experiments\cite{Kapner:2006si}. When $\delta \geq
3$, where $\rm R < 1~{\rm nm}$, it is possible to detect a graviton signal
at high energy colliders.

At colliders, the exchange of a virtual KK graviton or the emission of a real
KK mode could give rise to interesting phenomenological signals at
TeV scale\cite{ADD:Gian,ADD:HanTao}. The virtual effects of the KK modes
could lead to the enhancement of the cross section of pair
productions in the processes, for example, dilepton, digauge boson
($\gamma\gamma$, $\rm ZZ$, $\rm W^+W^-$), dijet, $\rm t\bar{t}$ pair,
HH pair\cite{ADDvirtualA, ADDvirtualB, ADDvirtualC, ADDvirtualD,
ADDvirtualE, ADDvirtualF, ADDvirtualG}, etc. The real emission of a
KK mode could lead to large missing $\rm E_T$ signals viz. mono jet,
mono gauge boson\cite{ADD:Gian, ADD:HanTao, ADDrealA, ADDrealB}, etc.
The CMS Collaboration has performed a lot of research for LED on
different final states at $\rm \sqrt{s}=7$
TeV\cite{CMS:LED1,CMS:LED2,CMS:LED3}, and they set the most
stringent lower limits to date to be $\rm 2.5~{\rm TeV} < M_S <3.8~{\rm TeV}$
by combining the diphoton, dimuon and dielectron channels.

\begin{figure}[hbtp]
\vspace{-5cm}
\hspace*{-3cm}
\centering
\includegraphics[scale=1]{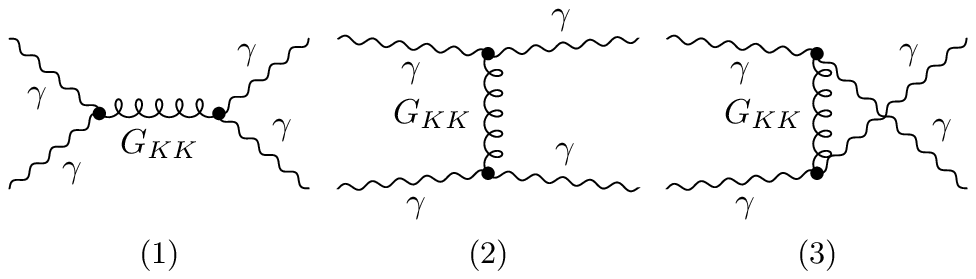}
\vspace{-22cm}
\caption{\label{figrrGrr} Feynman diagrams for light-light scattering of the diphoton production
through graviton exchange in the LED model.}
\end{figure}

Now we give the analytical calculations of the process $\gamma\gamma\rightarrow \gamma\gamma$
at the LHC in the LED model. In our calculation we use the de Donder gauge.
The relevant Feynman rules involving a graviton in the LED model can be found in Ref.\cite{ADD:HanTao}.
We denote the process as
\begin{eqnarray}
\gamma(p_1) + \gamma(p_2) \rightarrow \gamma(p_3) + \gamma(p_4)
\end{eqnarray}
where $\rm p_1$, $\rm p_2$ and $\rm p_3$, $\rm p_4$ represent the momenta of
the incoming and outgoing particles respectively.
In Fig.\ref{figrrGrr} we display the Feynman diagrams for this
process in the LED model. We see the diphoton productions going through
s-, t-, u-channels of graviton exchange.
The spin-0 states only couple through the dilaton
modes, which have no contribution to $\gamma\gamma$ processes.
So we focus our study on the spin-2 component of the KK states.
The couplings between gravitons and SM particles are proportional to
a constant named gravitational coupling $\rm \kappa \equiv \sqrt{16 \pi G_N}$,
which can be expressed in terms of the fundamental Planck scale $\rm M_S$
and the size of the compactified space R by
\begin{eqnarray}
\rm  \kappa^2 R^{\delta} = \rm 8 \pi (4 \pi)^{\delta/2} \Gamma(\delta/2) M_S^{-(\delta+2)}
\end{eqnarray}
In practical experiments, the contributions of the different
Kaluza-Klein modes have to be summed up, so the propagator is
proportional to $\rm i/(s_{ij}-m^2_{\vec{n}})$, where
$\rm s_{ij}=(p_i+p_j)^2$ and $\rm m_{\vec{n}}$ is the mass of the KK state
$\rm \vec{n}$. Thus, when the effects of all the KK states are taken
together, the amplitude is proportional to $\rm \sum\limits_{\vec{n}}
\frac{i}{s_{ij}-m^2_{\vec{n}}+i\epsilon}=D(s)$. If $\delta \geq 2$
this summation is formally divergent as $\rm m_{\vec{n}}$ becomes large.
We assume that the distribution has a ultraviolet cutoff at
$\rm m_{\vec{n}}\sim M_S$, where the underlying theory becomes manifest.
Then $\rm D(s)$ can be expressed as
\begin{eqnarray}
\rm  D(s) =\rm \frac{1}{\kappa^2}
\frac{8\pi}{M_S^4}(\frac{\sqrt{s}}{M_S})^{\delta-2}[\pi + 2i
I(M_S/\sqrt{s})].
\end{eqnarray}
The imaginary part I($\Lambda/\sqrt{s}$) is from the summation over
the many non-resonant KK states and  its expression can be found in
Ref.\cite{ADD:HanTao}. Finally the KK graviton propagator after
summing over the KK states is
\begin{eqnarray}
\label{prop}
\rm \tilde{G}^{\mu\nu\alpha\beta}_{KK}=D(s)
 \left(\eta_{\mu\alpha}\eta_{\nu\beta} + \eta_{\mu\beta}\eta_{\nu \alpha}
- \frac{2}{D-2}\eta_{\mu\nu}\eta_{\alpha\beta}\right).
\end{eqnarray}
Using the Feynman rules in the LED model and the propagator given by
Eq.(\ref{prop}), we can get the amplitudes for the virtual KK
graviton exchange diagrams in Fig.\ref{figrrGrr}.
For the formulas and details we refer to the two original calculations\cite{rr2rr1,rr2rr2}
for more details. Inserting the amplitudes into Eqs.(\ref{totcrosssection})
and (\ref{partoiccrosssection}) we can obtain the total cross section.

\section{Numerical Results}

\subsection{Input parameters and Kinematic Cuts}
We take the input parameters as $\rm M_p=0.938272046\ GeV$,
$\rm \alpha_{ew}(m^2_Z)^{-1}|_{\overline{MS}}=127.918$,
$\rm m_Z=91.1876 GeV$, $\rm m_W=80.385\ GeV$\cite{2012PDG}
and we have $\rm sin^2\theta_W=1-(m_W/m_Z)^2=0.222897$.
Light fermion masses are needed in the SM one-loop $\gamma\gamma$ production
and we take $\rm m_e=.510998910\ MeV$, $\rm m_\mu=105.658367\ MeV$,
$\rm m_\tau=1776.82\ MeV$, $\rm m_u=m_d=53.8\ MeV$, $\rm m_c=1.27\ GeV$,
$\rm m_s=101\ MeV$, $\rm m_b=4.67\ GeV$.
The top quark pole mass is set to be $\rm m_t=173.5\ GeV$.
The colliding energy in the proton-proton c.m.s. system
is assumed to be $\rm \sqrt{s}=14\ TeV$ at future LHC.
We use FeynArts, FormCalc and LoopTools (FFL)\cite{FeynArts,FormCalc,LoopTools}
packages to generate the amplitudes performing the numerical calculations.
We adopt BASES\cite{BASES} to do the phase space integration.
Based on the forward proton detectors to be installed by the CMS-TOTEM and
the ATLAS collaborations we choose the detected acceptances to be
\begin{itemize}
 \item CMS-TOTEM forward detectors with $0.0015<\xi_1<0.5$
 \item CMS-TOTEM forward detectors with $0.1<\xi_2<0.5$
 \item AFP-ATLAS forward detectors with $0.0015<\xi_3<0.15$
\end{itemize}
which we simply refer to as $\xi_1$, $\xi_2$ and $\xi_3$, respectively.
During the calculation we use $\xi_1$ in default unless otherwise stated.
The survival probability in the $\gamma \gamma$ collision is
taken to be 0.9 for central diphoton exchange and 0.03 for DPE processes\cite{SurvivalP}.

In all the results present in our work, we impose a cut of pseudorapidity $|\eta| < 2.5$
for the final state particles since central detectors of the ATLAS and CMS have a pseudorapidity
$|\eta|$ coverage of 2.5. The general acceptance cuts for all the signal and background events are
\begin{eqnarray} \label{cuts} \nonumber
\rm \omega>300\ GeV, p_T^{\gamma}>20\ GeV, |\eta|^{\gamma}<2.5, \Delta R(\gamma\gamma)>0.4
\end{eqnarray}
where $\rm \Delta R = \sqrt{\Delta \Phi^2 + \Delta \eta^2}$ is the separation
in the rapidity-azimuth plane and $\rm p_T$ is the transverse momentum of the photons.

\subsection{Background Analysis}

Our signal topology is simply that of two photons in the final states excited by graviton exchange.
The main background comes from two sources: one is the background coming from SM
$\rm pp\rightarrow p\gamma\gamma p\rightarrow p\gamma\gamma p$ production through one-loop contributions.
The other comes from diffractive Double Pomeron Exchange (DPE)
[or Central Exclusive Diffractive (CED) production] of the diphoton.

Let us first consider the SM $\gamma\gamma\rightarrow \gamma\gamma$ one-loop production.
Though there is no tree level contribution to $\gamma\gamma\rightarrow\gamma\gamma$,
it is still important to include loop effects of its contributions.
The precise determination of this cross section has a special importance
to test the renormalization procedure of the parts of the SM containing W gauge bosons.
Moreover, as in our case, this loop process becomes
a background for new physics searches through $\gamma\gamma$ scattering.
One-loop diagrams involve charged fermions and W bosons rotate in loop.
Here we include all these contributions and perform the calculation with the FFL package.
The calculation can also be found in Ref.\cite{EDprrp2}.

Now let us see the background contributions from DPE or CED productions.
Different from Ref.\cite{EDprrp2} where one takes
the SM $\gamma\gamma$ production as the only background,
here we also consider the DPE backgrounds. However, their results are for
$0.1<\xi_2<0.5$, where in this range, DPE indeed is small and not very important.
But for the other choice of the forward detector acceptance it might be
interesting to study them.
In DPE, two colorless objects are emitted from both protons.
Their partonic components are resolved and create a heavy mass object X
in the central detector in the Pomeron-Pomeron interaction.
The event is characterized by two rapidity gaps between the
central object and the protons.
Through the exchange of two Pomerons a dijet system, a diphoton system, a WW and
ZZ pair, or a Drell-Yan pair can be created for instance.
The DPE production cross section can be obtained within the factorized Ingelman-Schlein\cite{ISmodel}.
In the Ingelman-Schlein model, one assumes that the Pomeron has a well-defined partonic structure,
and that the hard process takes place in a Pomeron-proton or proton-Pomeron (single diffraction)
or Pomeron-Pomeron (central diffraction) processes.
These processes are described in terms of the proton "diffractive" parton distribution functions (DPDFs).
Similar to the parton distribution functions,
the DPDFs are also obtained from Deep Inelastic Scattering experiments\cite{H1DPDF1,H1DPDF2}.
The difference is that they contain a dependence on additional variables describing the proton kinematics:
the relative energy loss $\xi$ and four-momentum transfer t, which we introduce as
\begin{eqnarray}\nonumber
\rm g^D(x,\mu^2) &=&\rm \int dx_{\textbf{P}} d\beta \delta(x-x_{\textbf{P}}\beta)
g_{\textbf{P}}(\beta, \mu^2) f_{\textbf{P}}(x_{\textbf{P}})    \\
&=&\rm  \int^1_x \frac{dx_{\textbf{P}}}{x_{\textbf{P}}} f_{\textbf{P}}(x_{\textbf{P}})
g_{\textbf{P}}(\frac{x}{x_{\textbf{P}}},\mu^2)
\end{eqnarray}
where $\rm g^D$ denotes either the quark or the gluon distributions with D refers to "diffractive".
$\rm f_{\textbf{P}}(x_{\textbf{P}})$ is the flux of the Pomerons and is expressed as
\begin{eqnarray}
\rm f_{\textbf{P}}(x_{\textbf{P}})=\int^{t_{max}}_{t_{min}} f(x_{\textbf{P}},t)  dt
\end{eqnarray}
with $\rm t_{min}$ and $\rm t_{max}$ being kinematic boundaries.
$\rm g_{\textbf{P}}(\beta,\mu^2)$ is the partonic structure of the Pomeron.
$\rm x_{\textbf{P}}$ here is the fraction of the proton momentum carried by the Pomeron corresponding
to the relative energy loss $\xi$ and $\beta$ is the fraction of the Pomeron momentum carried by the struck parton.
Both Pomeron flux factors $\rm f(x_{\textbf{P}},t)$ as well as quark and gluon distributions
in the Pomeron $\rm g_{\textbf{P}} (\beta, \mu^2)$ were taken
from the H1 collaboration analysis of the diffractive structure function
at HERA\cite{H1DPDF1,H1DPDF2}.
Therefore, the final forward detector acceptance's $\xi$-dependent
convolution integral for the DPE production is given by
\begin{eqnarray}\label{Eq.DPE} \nonumber
\rm \sigma &=& \rm \sum_{ij=q\bar{q},gg}  \int^{\xi_{max}}_{\frac{M_{inv}}{\sqrt{s}}} 2z dz \int^{Min(\xi_{max},
\frac{z^2}{\xi_{min}})}_{Max(\frac{z^2}{\xi_{max}},\xi_{min})} \frac{dx_1}{x_1}
\int^1_{x_1} \frac{dx_{\textbf{P}_i}}{x_{\textbf{P}_i}} f_{\textbf{P}_i}(x_{\textbf{P}_i})
g_{\textbf{P}_i}(\frac{x_1}{x_{\textbf{P}_i}},\mu^2) \\
&& \rm  \int^1_{\frac{z^2}{x_1}} \frac{dx_{\textbf{P}_j}}{x_{\textbf{P}_j}} f_{\textbf{P}_j}(x_{\textbf{P}_j})
g_{\textbf{P}_j}(\frac{z^2}{x_1 x_{\textbf{P}_j}},\mu^2) \hat\sigma(ij\rightarrow \gamma\gamma,\hat{s}=z^2 s,\mu) .
\end{eqnarray}
Diffraction usually dominates for $\xi<0.05$ which
means that we can replace $\rm Min(\xi_{max}, \frac{z^2}{\xi_{min}})$ by 0.05 directly.
We check this and find the difference is quite small, as expected.
The DPE production cross section should be multiplied by the gap survival probability
S=0.03 for LHC. This is different for $\gamma\gamma$ exchange where S=0.9
and single diffractive production with S=0.6\cite{SurvivalP}.

\begin{figure}[hbtp]
\vspace{-5cm}
\hspace*{-3cm}
\centering
\includegraphics[scale=1]{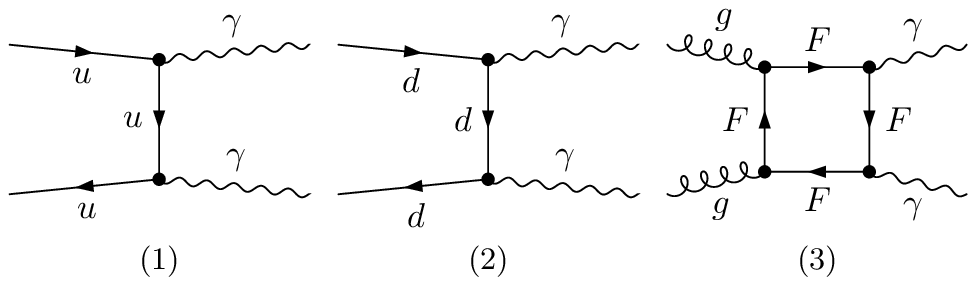}
\vspace{-22cm}
\caption{\label{figDPE} Feynman diagrams for DPE contributions in $\rm u\bar{u}$, $\rm d\bar{d}$
and gg collision modes.}
\end{figure}

We separate the DPE background contributions into three different parts
based on different parton-collision modes include $\rm u\bar{u}$, $\rm d\bar{d}$
and gg. Part of the Feynman diagrams are presented in Fig.\ref{figDPE}.
Other diagrams include the change of the loop arrow and cross change
of the legs, which are similar and thus are not shown. F means the class of fermions.
$\rm u\bar{u}$ and $\rm d\bar{d}$ contributions are considered at tree level.
Typically, we also include the gg collision mode contribution where there are no tree level diagrams
but we start from gluon-gluon fusion at one-loop level through fermion loops.
We do this because we want to see how large their contributions will be,
especially the gg collision modes, since usually one does not consider them.
Indeed we find that the gg contribution is comparable with the quark-quark collision mode
even though they contribute through loops. See our discussion below.

\begin{figure}[hbtp]
\centering
\includegraphics[scale=0.6]{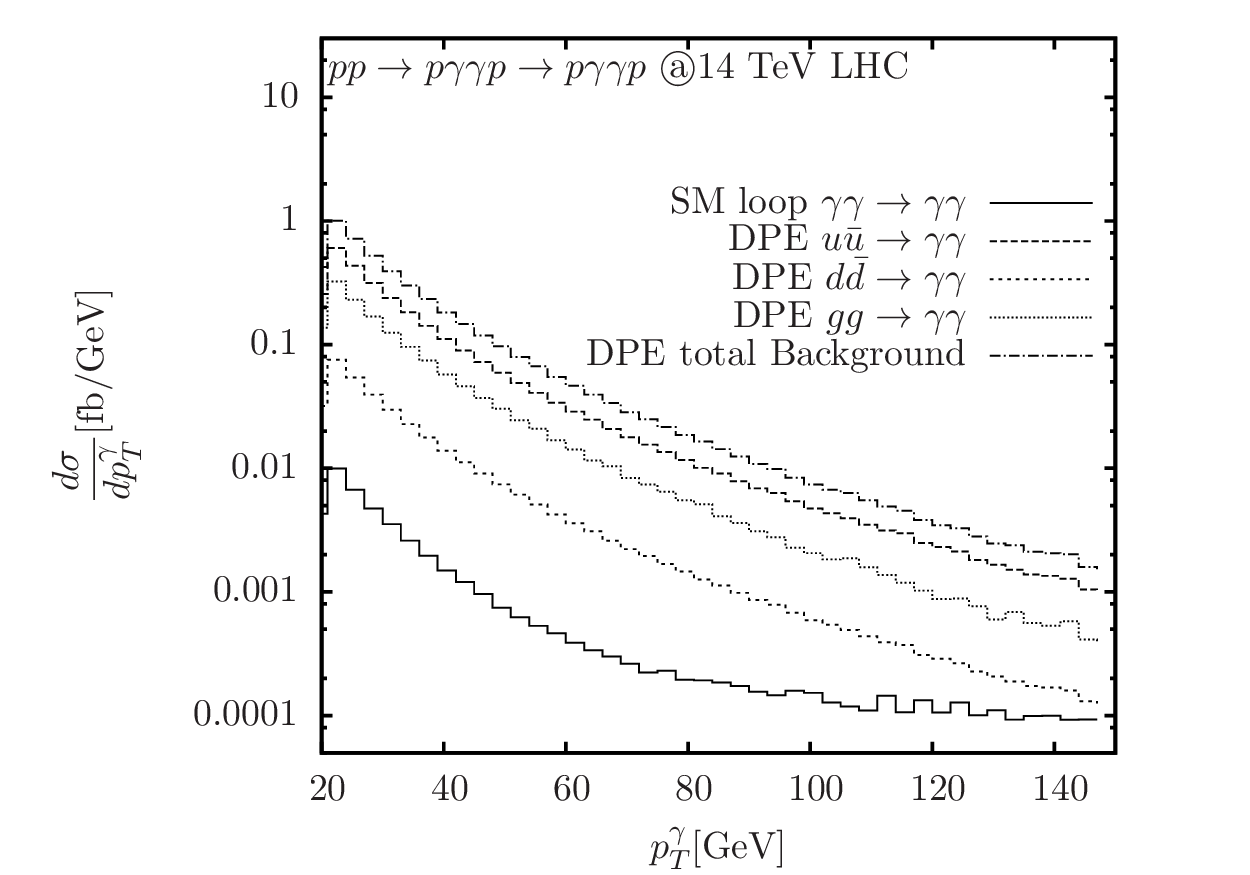}
\includegraphics[scale=0.6]{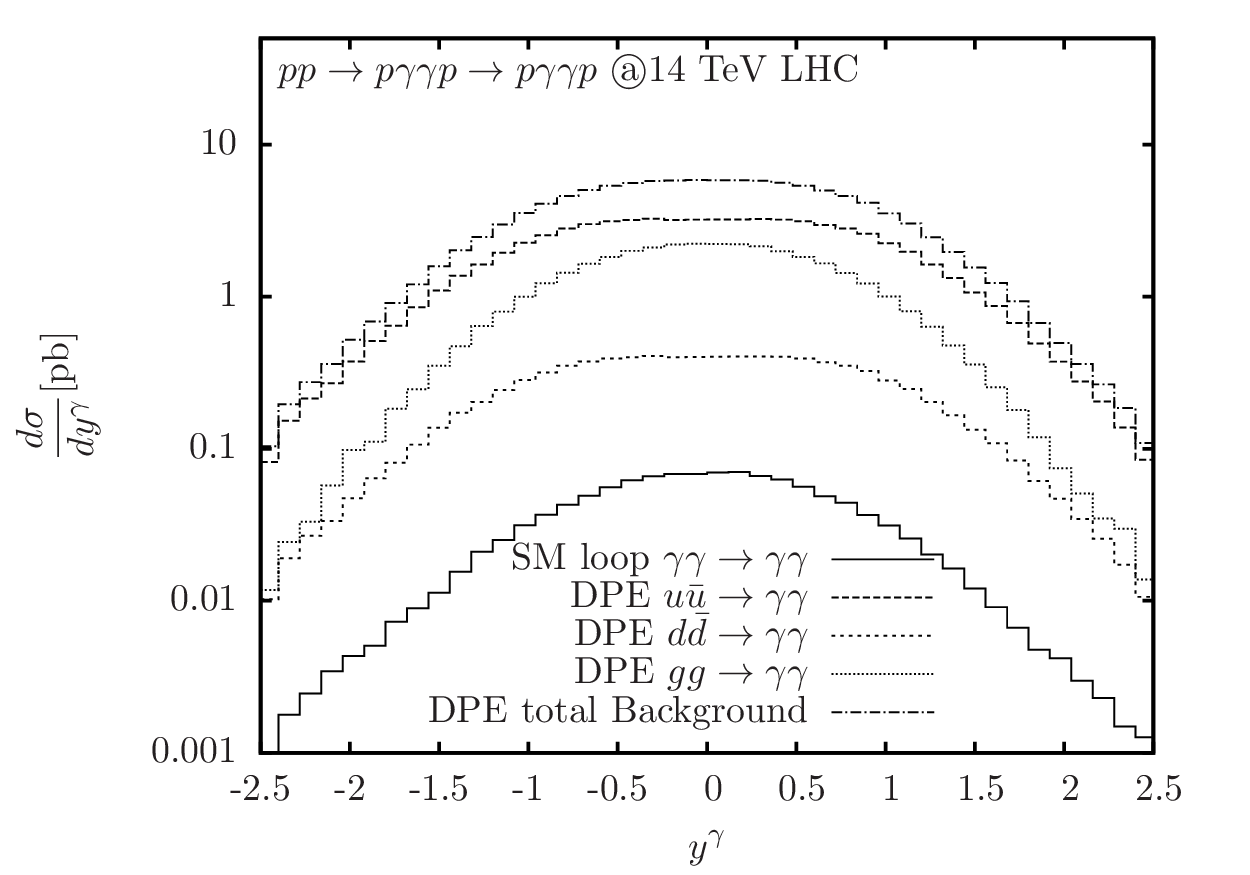}
\caption{\label{BGdis}
The transverse momentum ($\rm p_T$) and
Rapidity (y) distribution of a leading photon for the background production of
$\rm pp\rightarrow p\gamma\gamma p\rightarrow p\gamma\gamma p$,
on the basis of their transverse momentum $\rm p_T^{\gamma_1}\geq p_T^{\gamma_2}$.
The kinematic cuts are considered and no survival probability factor is taken into account.
Solid curve present the SM $\gamma\gamma$ loop contribution.
Dashed, short-dashed, dotted and dashed-dotted lines refer to
the DPE $\rm u\bar{u}$-collision mode, $\rm d\bar{d}$-collision mode, gg-collision mode
and their sum, respectively. Here $0.0015<\xi_1<0.5$.}
\end{figure}

Before selecting our event samples for diphoton production,
we order the two photons on the basis of their transverse momenta, i.e.,
$\rm p_T^{\gamma_1}\geq p_T^{\gamma_2}$. $\rm p_T^{\gamma_1}$ is named the leading photon and
simply is denoted as $\rm p_T^{\gamma}$ in the following text and figures.
Figure \ref{BGdis} presents the diphoton background productions.
Different parton collision modes are considered in order to clearly compare
their contributions. At this moment, no survival probability factor is taken into account.
Dashed, short-dashed, dotted and dashed-dotted lines refer to the
DPE $\rm u\bar{u}$-collision mode, $\rm d\bar{d}$-collision mode, gg-collision mode
and their sum, respectively.
Here u include both u-quark and c-quark and d refer to d, s and b-quarks.
We can see the $\rm u\bar{u}$-collision mode is the largest one.
gg-collision mode, though there is no tree level contribution and
it only appears at the loop level, its contribution is comparable with the quark collision.
Both are larger than the $\rm d\bar{d}$ collision mode.
When no survival probability factor is considered, there contributions are much larger than
the SM $\gamma\gamma$ loop contributions. But they become comparable when
the survival probability factor is taken into account.
More details can be found in Table \ref{BGsigma}.
After considering these, the survival probability factor
contributions of DPE are only three times over the SM loop contribution.
The total backgrounds before and after taking into account the survival probability factor
are 14.2846 and 0.5516 fb, respectively.

\vspace*{0.4cm}
\begin{table}[hbtp]
\begin{center}
\begin{tabular}{c c c c c c}	
\hline\hline
$\sigma$[fb] & \multicolumn{5}{c}{ Background contribution for diphoton production} \\ [0.5ex]
$\rm (S_{DPE},S_{\gamma\gamma})$  &$\rm u\bar{u}$&$\rm d\bar{d}$&gg&$\rm SM_{loop}$ &total\\
\hline
(1,1)     & 8.5792    &1.0724     &4.49159  &0.1415  &14.2846 \\
(0.03,0.9)& 0.2574    &0.0322     &0.1347   &0.1273  &0.5516\\
\hline\hline
\end{tabular}
\end{center}
\vspace*{-0.8cm}
\begin{center}
\begin{minipage}{14cm}
\caption{\label{BGsigma} Background contribution for diphoton production
in different parton collision modes before and after taking into account the
survival probability factor.}
\end{minipage}
\end{center}
\end{table}

\subsection{Signal boundary at future LHC}

\begin{figure}[hbtp]
\centering
\includegraphics[scale=0.8]{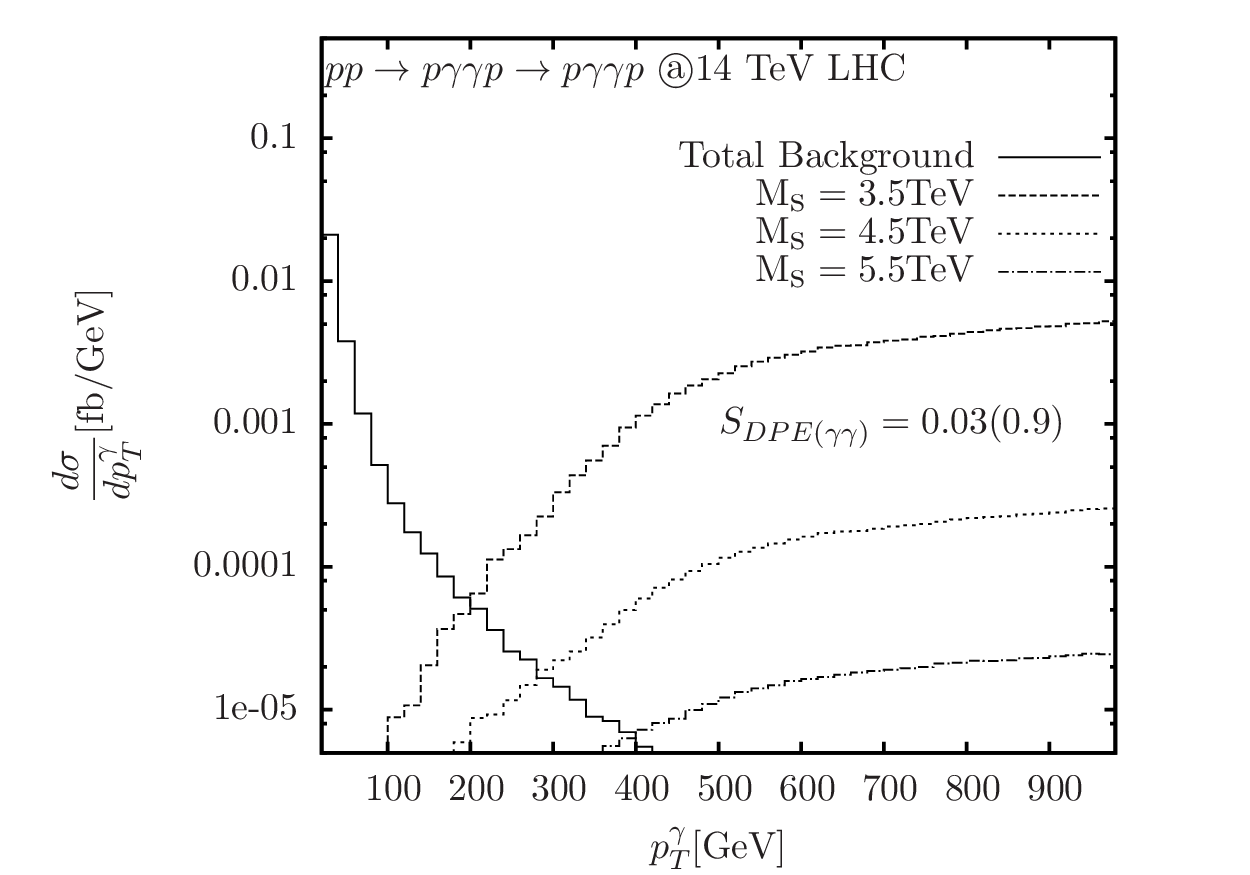}
\caption{\label{pt1000}
The signal and total background transverse momentum ($\rm p_T^{\gamma}$)
distribution of leading photon for the process
$\rm pp\rightarrow p\gamma\gamma p\rightarrow p\gamma\gamma p$, on the basis of their
transverse momentum $\rm p_T^{\gamma_1}\geq p_T^{\gamma_2}$ with $\rm M_S = 3.5, 4.5, 5.5 TeV$,
$\rm \sqrt{s}=14 TeV$ and $\delta=4$. The high transverse momentum range includes up to 1 TeV.
The kinematic cuts and survival probability factor $\rm S_{DPE}=0.03$
and $\rm S_{\gamma\gamma}=0.9$ are taken into account.}
\end{figure}

We are interested in the $\rm p_{T}^{\gamma}$ distributions which are displayed in Fig.\ref{pt1000}.
Both the transverse momentum of signal and the total background
distribution of leading photon are presented.
The range of the transverse momentum is displayed up to 1 TeV.
The kinematic cuts and survival probability factor $\rm S_{DPE}=0.03$
and $\rm S_{\gamma\gamma}=0.9$ are taken into account.
The solid line refers to the total background including the sum of the SM $\gamma\gamma\rightarrow \gamma\gamma$
loop contribution and the DPE diphoton contributions.
It is interesting that most of their contributions are in the low $\rm p_T$ range.
One may observe that, with a $\rm p_T^{\gamma}$ cut larger than $\sim$300 GeV, 
all the background contributions can be
reduced, while leaving the signal changes slight.
Using this feature the light-by-light scattering of the diphoton central production at the LHC can
be almost background free and can be detected precisely.
If, in this case, a new physics signal appears mainly in the high transverse momentum range,
light-by-light scattering can be very sensitive to it.
This is exactly the LED effects we studied here.
In this case, in the high $\rm p_{T}^{\gamma}$ region, the LED effects dominate the signal
distribution since more KK modes contribute with the increase of $\rm p_T$.
We use dashed, dotted and dashed-dotted lines to refer to the LED signal
with $\rm M_S$ equal 3.5, 4.5 and 5.5 TeV, respectively.
Just to note, in the following data analysis, we do not take the cut $\rm p_T^{\gamma}>300GeV$.
Nevertheless, we can still use diphoton production to test
the LED effects up to high energy scales.

\begin{figure}[hbtp]
\centering
\includegraphics[scale=0.8]{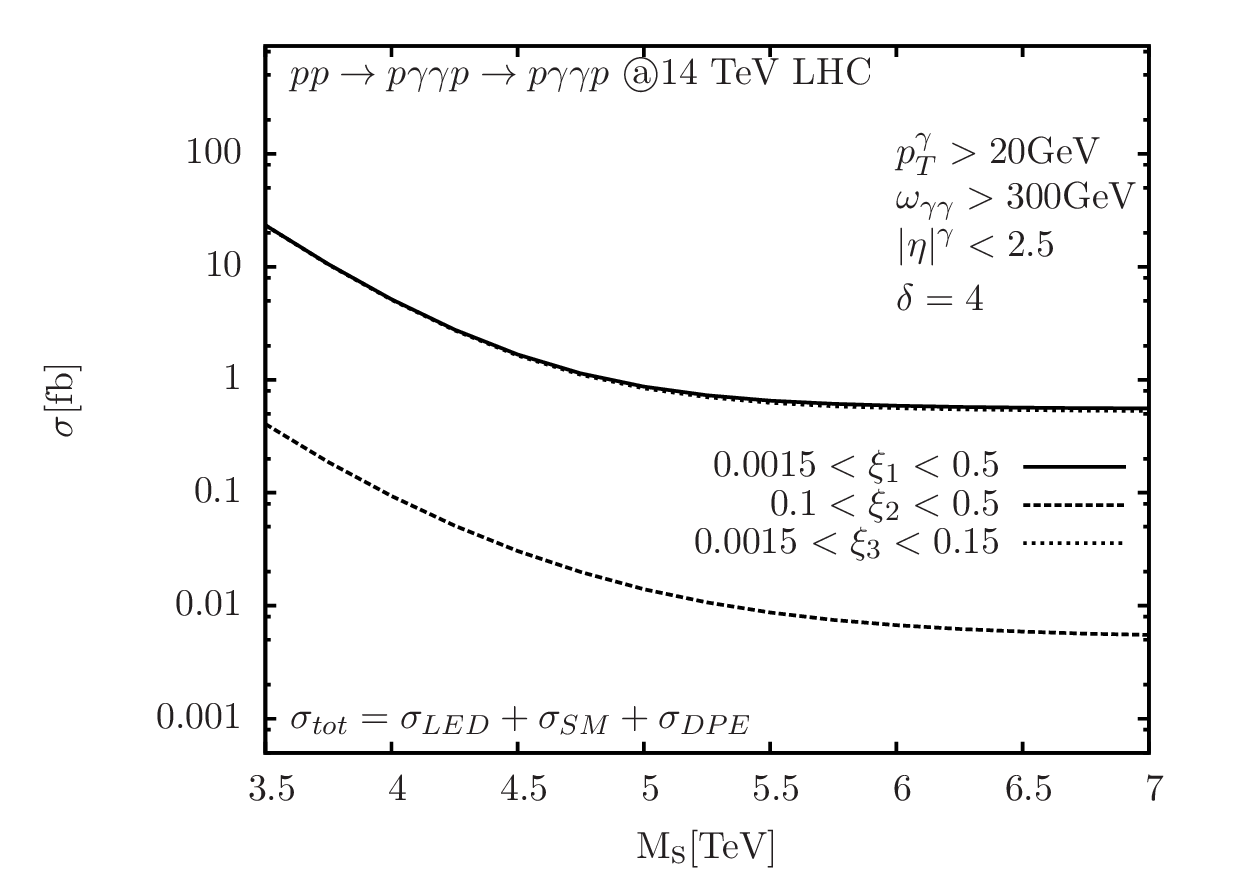}
\caption{\label{mssigma} The total [signal plus background] cross sections for the process
$\rm pp\rightarrow p\gamma\gamma p\rightarrow p\gamma\gamma p$ in the SM and LED model
as the function of $\rm M_S$ with $0.0015<\xi_1<0.5$, $0.1<\xi_2<0.5$, $0.0015<\xi_3<0.15$
and $\delta=4$. The kinematic cuts and survival probability factor $\rm S_{DPE}=0.03$
and $\rm S_{\gamma\gamma}=0.9$ are taken into account.}
\end{figure}

In Fig.\ref{mssigma} we present the dependence of the cross section on
the energy scale $\rm M_S$ with $0.0015<\xi_1<0.5$, $0.1<\xi_2<0.5$ and $0.0015<\xi_3<0.15$
correspond to solid, dashed and dotted lines, respectively.
We present the curves for the cross sections with the extra dimension $\delta=4$.
The curves present the signal plus its corresponding background.
The kinematic cuts in Eq.(\ref{cuts}) are considered and the survival probability factor $\rm S_{DPE}=0.03$
and $\rm S_{\gamma\gamma}=0.9$ for central production at the LHC are taken into account.
It is clear that, for a given value of $\delta$, the cross section
decreases rapidly with the increment of $\rm M_S$, and finally approaches its corresponding background results.
We can see that the cross sections of $\xi_1$ and $\xi_3$ are very close to each other,
and both are larger than the results of $\xi_3$, two orders of magnitude larger, both for the signal
and the background.

\begin{figure}[hbtp]
\centering
\includegraphics[scale=0.7]{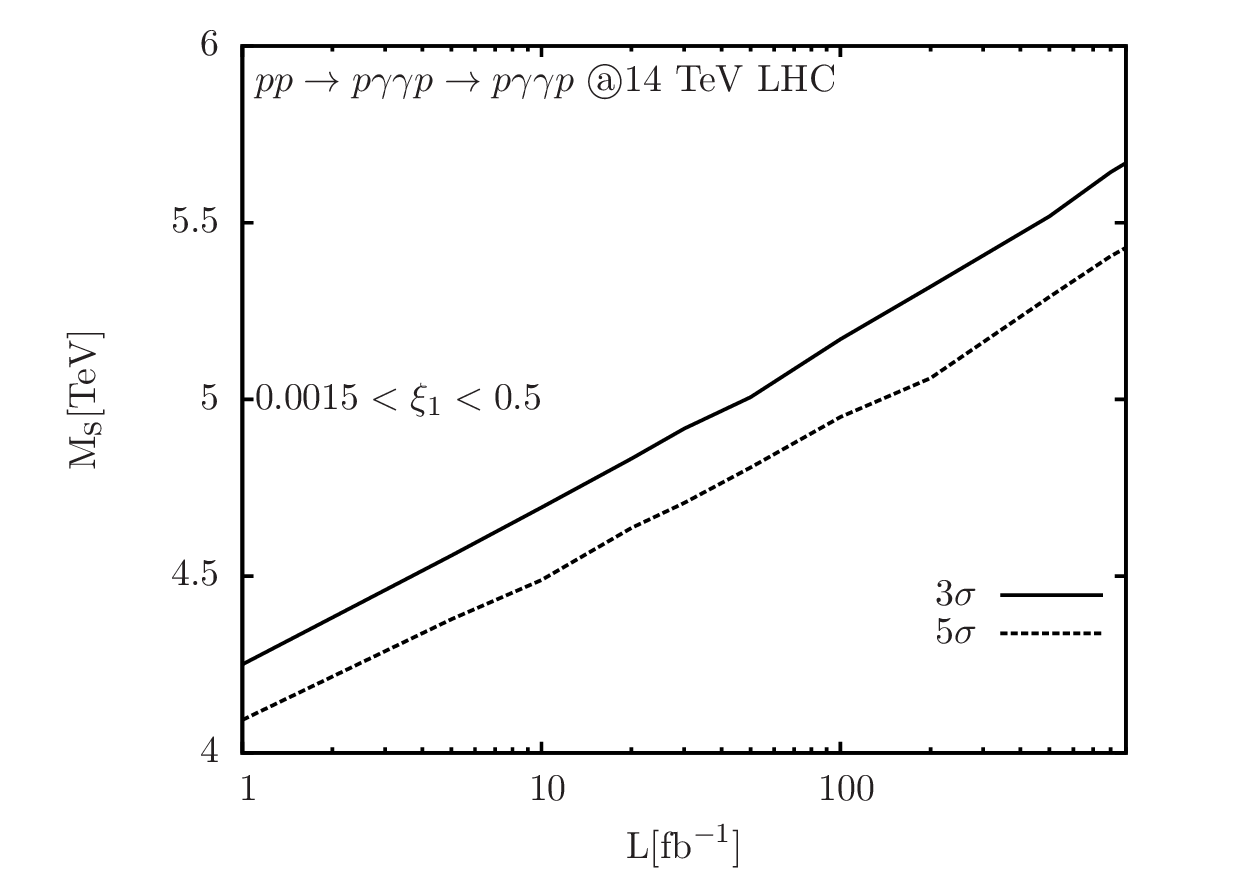}
\includegraphics[scale=0.7]{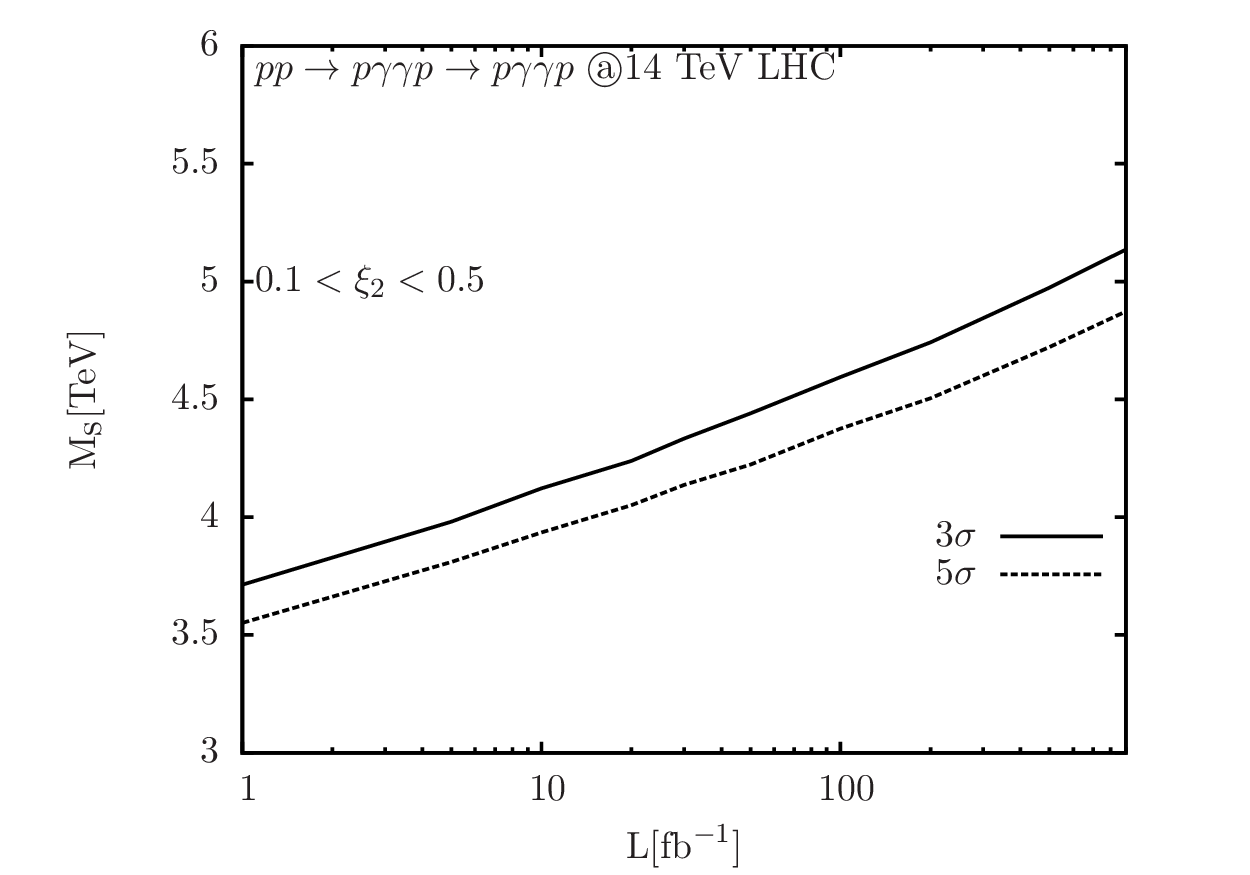}
\includegraphics[scale=0.7]{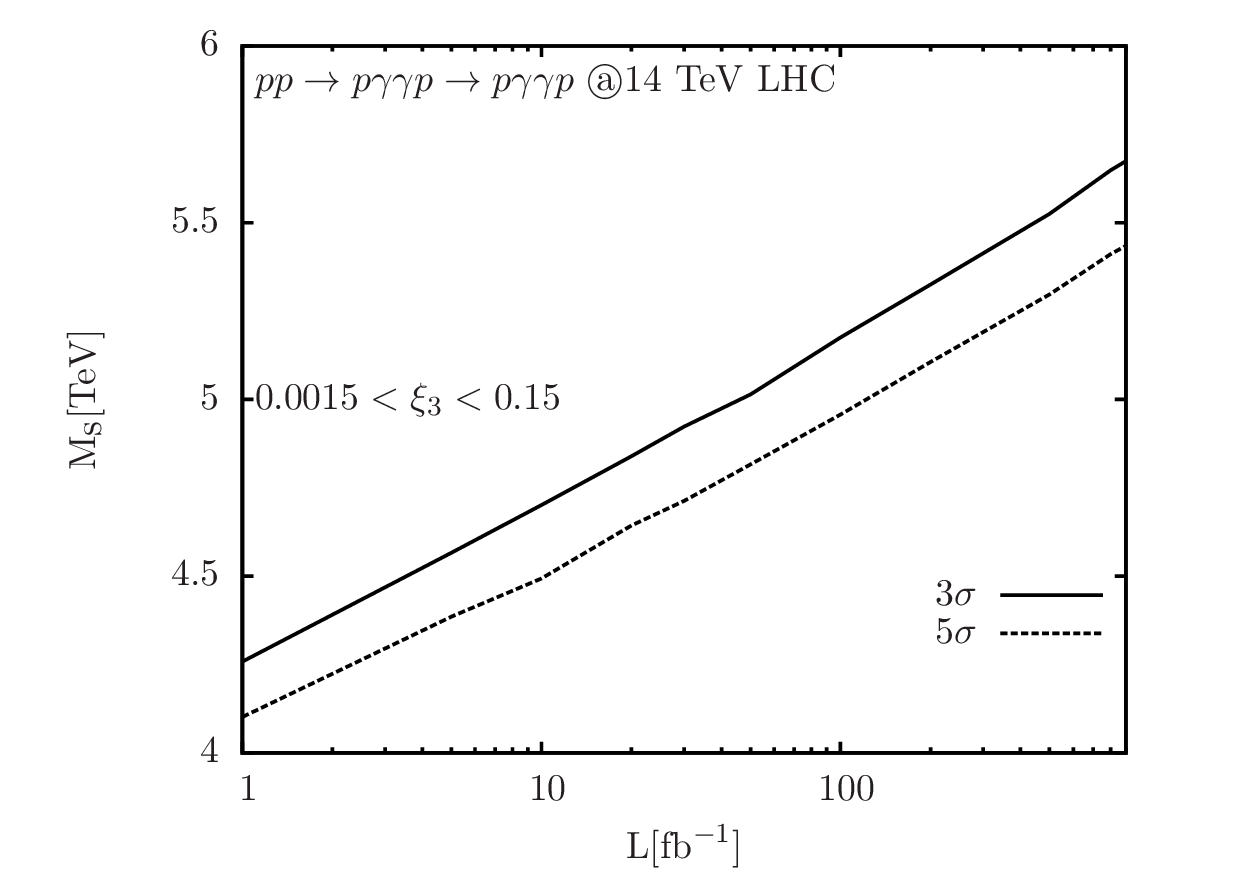}
\caption{\label{msbound}
The $\rm M_S$ bounds for $3\sigma$ [solid lines] and $5\sigma$ [dashed lines] deviations
from the total backgrounds as functions of the luminosity ${\cal L}$ with extra dimensions
$\delta=4$ for different choice of forward detector acceptance $\xi$.}
\end{figure}

To analyze the potential of the $\gamma\gamma$-collision
at the high energy LHC to probe the diphoton signal through LED effects,
we define the statistical significance (SS) of the signal:
\begin{eqnarray}
\rm SS=\frac{|\sigma_{S+B}-\sigma_{B}|}{\sqrt{\sigma_B}} \sqrt{\cal L}
\end{eqnarray}
where ${\cal L}$ is the integrated luminosity of the $\gamma\gamma$ option of the LHC.
In the $\rm {\cal L}-M_S$ parameter space, we present the plots for the $3\sigma$
and $5\sigma$ deviations of the signal from the total backgrounds.
We display our results for the forward detector acceptance chosen
to be $0.0015<\xi_1<0.5$, $0.1<\xi_2<0.5$ and $0.0015<\xi_3<0.15$
in Fig.\ref{msbound}; see the first, the second and the third panel, respectively.
The c.m.s. energies of the LHC collider is $\sqrt{s}$ = 14 TeV.
The extra dimension is set to be $\delta=4$.
Concerning the criteria, for a statistical significance of $\rm SS > 5$,
with an integrated luminosity $\rm {\cal L} = 100 fb^{-1}$,
$\rm pp\rightarrow p\gamma\gamma p\rightarrow p\gamma\gamma p$
production through graviton exchange can probe the LED effects
up to $\rm M_S=4.95 (4.38, 4.96) TeV$ for $\xi_1 (\xi_2, \xi_3)$.
With an integrated luminosity $\rm {\cal L} = 200 fb^{-1}$
one can probe the LED effects up to the scale
$\rm M_S=5.06 (4.51, 5.11) TeV$ for $\xi_1 (\xi_2, \xi_3)$, respectively.
The current limits on $\rm M_S$ of the large extra dimensions
are about 3 - 5 TeV from recent ATLAS and CMS results.
Our results show that the process we considered can probe
the LED effects only moderately above the current experimentally limits.
The limits we obtained are indeed worse than that obtained
by other conventional approaches like dijet production
signals at the LHC, which give a reach of about 10 TeV for $\rm M_S$.
Moreover, this is a little worse than or comparable to the diphoton production at the LHC which
can extend this search to 5.3-6.7 TeV and the Drell-Yan process
which can drive the search up to about 5.8 TeV.
However, the limits we obtained are better than the results obtained
by monojet signals (i.e., 4.5; 3.3 TeV for n = 2; 6),
or a similar photon-induced study at the LHC using the subprocess
$\gamma\gamma \rightarrow \ell^+\ell^-$\cite{EDpllp1}.

\section{Summary}

A lot of theoretical work has been done to study the LED effects.
Even if a process can be traced back to a definite set of operators,
it is rarely the case that a
particular collider signature can be traced back to a unique
process.  For this reason many different, complementary
measurements are usually required to uncover the underlying new physics processes.
Observing light-by-light scattering at the LHC
has recently received much more attention\cite{lightlight,lightlight1}.
It is believed to be a clean and sensitive channel to possible new physics.
In this paper, we calculate the diphoton production at the LHC
via the process $\rm pp\rightarrow p\gamma\gamma p\rightarrow p\gamma\gamma p$
in the SM and LED. Typically, when we do a background analysis, we also
study the DPE of $\gamma\gamma$ production.
We present their cross section dependence on the energy loss of the proton $\xi$
and compare its production separately in the quark-quark collision mode to the gluon-gluon collision mode.
We conclude that the contribution from the gluon-gluon collision mode are comparable to the
quark-quark collision mode. A $\rm p_T^\gamma > 300\ GeV$ cut can suppress the DPE
background efficiently. However, if there are no kinematic cuts taken into account,
especially in the low $\xi$ range, they should better be studied and considered.
Finally, we present the $\rm M_S$ bounds for the $3\sigma$ and $5\sigma$ deviations
from the total backgrounds as functions of the luminosity ${\cal L}$ with extra dimensions
$\delta=4$ for different choices of the forward detector acceptance $\xi$.
Concerning the criteria, for a statistical significance of $\rm SS > 5$,
with an integrated luminosity $\rm {\cal L} = 200 fb^{-1}$,
$\rm pp\rightarrow p\gamma\gamma p\rightarrow p\gamma\gamma p$
production through graviton exchange can probe the LED effects up to the scale
$\rm M_S=5.06(4.51,5.11) TeV$ for $\xi_1(\xi_2,\xi_3)	$, respectively, where
$0.0015<\xi_1<0.5$, $0.1<\xi_2<0.5$ and $0.0015<\xi_3<0.15$.

\section*{Acknowledgments} \hspace{5mm}
Project supported by the National Natural Science Foundation of China (No. 11205070),
Shandong Province Natural Science Foundation (No. ZR2012AQ017)
and by the Fundamental Research Funds for the Central Universities (No. DUT13RC(3)30).

\end{document}